\documentclass[twoside]{article}

\makeatletter
\@ifundefined{MPFourScale}{\def\MPFourScale{1.00}}{}
\DeclareFontFamily{U}{mp4}{}%
\DeclareFontShape{U}{mp4}{m}{n}{<->s * [\MPFourScale]cmb10}{}
\DeclareSymbolFont{boldgreekuc}{U}{mp4}{m}{n}
\DeclareMathSymbol{\bfAlpha}{\mathord}{boldgreekuc}{"41}
\DeclareMathSymbol{\bfBeta}{\mathord}{boldgreekuc}{"42}
\DeclareMathSymbol{\bfPsi}{\mathord}{boldgreekuc}{"09}
\DeclareMathSymbol{\bfDelta}{\mathord}{boldgreekuc}{"01}
\DeclareMathSymbol{\bfEpsilon}{\mathord}{boldgreekuc}{"45}
\DeclareMathSymbol{\bfPhi}{\mathord}{boldgreekuc}{"08}
\DeclareMathSymbol{\bfGamma}{\mathord}{boldgreekuc}{"00}
\DeclareMathSymbol{\bfEta}{\mathord}{boldgreekuc}{"48}
\DeclareMathSymbol{\bfIota}{\mathord}{boldgreekuc}{"49}
\DeclareMathSymbol{\bfXi}{\mathord}{boldgreekuc}{"04}
\DeclareMathSymbol{\bfKappa}{\mathord}{boldgreekuc}{"4B}
\DeclareMathSymbol{\bfLambda}{\mathord}{boldgreekuc}{"03}
\DeclareMathSymbol{\bfMu}{\mathord}{boldgreekuc}{"4D}
\DeclareMathSymbol{\bfNu}{\mathord}{boldgreekuc}{"4E}
\DeclareMathSymbol{\bfPi}{\mathord}{boldgreekuc}{"05}
\DeclareMathSymbol{\bfTheta}{\mathord}{boldgreekuc}{"02}
\DeclareMathSymbol{\bfRho}{\mathord}{boldgreekuc}{"52}
\DeclareMathSymbol{\bfSigma}{\mathord}{boldgreekuc}{"06}
\DeclareMathSymbol{\bfTau}{\mathord}{boldgreekuc}{"54}
\DeclareMathSymbol{\bfVartheta}{\mathord}{boldgreekuc}{"02} 
\DeclareMathSymbol{\bfOmega}{\mathord}{boldgreekuc}{"0A}
\DeclareMathSymbol{\bfVarphi}{\mathord}{boldgreekuc}{"08} 
\DeclareMathSymbol{\bfUpsilon}{\mathord}{boldgreekuc}{"07}
\DeclareMathSymbol{\bfZeta}{\mathord}{boldgreekuc}{"5A}

\DeclareFontFamily{U}{mp4sl}{}%
\DeclareFontShape{U}{mp4sl}{m}{n}{<->s * [\MPFourScale]cmmib10}{}
\DeclareSymbolFont{boldgreek}{U}{mp4sl}{m}{n}
\DeclareMathSymbol{\bfalpha}{\mathord}{boldgreek}{"0B}
\DeclareMathSymbol{\bfbeta}{\mathord}{boldgreek}{"0C}
\DeclareMathSymbol{\bfpsi}{\mathord}{boldgreek}{"20}
\DeclareMathSymbol{\bfdelta}{\mathord}{boldgreek}{"0E}
\DeclareMathSymbol{\bfepsilon}{\mathord}{boldgreek}{"0F}
\DeclareMathSymbol{\bfphi}{\mathord}{boldgreek}{"1E}
\DeclareMathSymbol{\bfgamma}{\mathord}{boldgreek}{"0D}
\DeclareMathSymbol{\bfeta}{\mathord}{boldgreek}{"11}
\DeclareMathSymbol{\bfiota}{\mathord}{boldgreek}{"13}
\DeclareMathSymbol{\bfxi}{\mathord}{boldgreek}{"18}
\DeclareMathSymbol{\bfkappa}{\mathord}{boldgreek}{"14}
\DeclareMathSymbol{\bflambda}{\mathord}{boldgreek}{"15}
\DeclareMathSymbol{\bfmu}{\mathord}{boldgreek}{"16}
\DeclareMathSymbol{\bfnu}{\mathord}{boldgreek}{"17}
\DeclareMathSymbol{\bfpi}{\mathord}{boldgreek}{"19}
\DeclareMathSymbol{\bfvartheta}{\mathord}{boldgreek}{"23}
\DeclareMathSymbol{\bfrho}{\mathord}{boldgreek}{"1A}
\DeclareMathSymbol{\bfsigma}{\mathord}{boldgreek}{"1B}
\DeclareMathSymbol{\bftau}{\mathord}{boldgreek}{"1C}
\DeclareMathSymbol{\bftheta}{\mathord}{boldgreek}{"12}
\DeclareMathSymbol{\bfomega}{\mathord}{boldgreek}{"21}
\DeclareMathSymbol{\bfvarphi}{\mathord}{boldgreek}{"27}
\DeclareMathSymbol{\bfchi}{\mathord}{boldgreek}{"1F}
\DeclareMathSymbol{\bfupsilon}{\mathord}{boldgreek}{"1D}
\DeclareMathSymbol{\bfzeta}{\mathord}{boldgreek}{"10}

\makeatother

\usepackage{amssymb}
\usepackage{graphicx}
\usepackage{epstopdf}
\newcommand{\bm}[1]{\mbox{\boldmath{$#1$}}}
\newcommand{\bdgr}[1]{\bm{#1}}

\newlength{\captionwidth}
\begin{document}
\thispagestyle{empty}
\begin{center}\vskip3pt
\textbf{\Large {D-Optimal Design for the Rasch Counts Model 
\\[2mm]
with Multiple Binary Predictors}}
\vspace{5mm}

  Ulrike Gra{\ss}hoff\,%
\renewcommand{\thefootnote}{\alph{footnote}}%
  \footnote{School of Business and Economics, Humboldt University Berlin, 
  Spandauer Stra{\ss}e 1, 10099 Berlin, Germany,
  \texttt{e-mail~grasshou@hu-berlin.de}},
  Heinz Holling%
\renewcommand{\thefootnote}{\fnsymbol{footnote}}%
\footnotemark[1]%
\renewcommand{\thefootnote}{\alph{footnote}}%
\footnote{Institute of Psychology, University of M\"unster, 
  Fliednerstra{\ss}e 21, 48149 M\"unster, Germany,
  \texttt{e-mail~holling@wwu.de}},
  Rainer Schwabe\,\footnote{University of Magdeburg, Institute for Mathematical Stochastics,
    Universit\"atsplatz 2, 39106 Magdeburg, Germany,
    \texttt{e-mail~rainer.schwabe@ovgu.de}}
\renewcommand{\thefootnote}{\fnsymbol{footnote}}%
    \footnotetext[1]{corresponding author}
\end{center}
\vspace{3mm}

\begin{abstract}
In this paper, we derive optimal designs for the Rasch Poisson counts model and the Rasch Poisson-Gamma counts model incorporating several binary predictors for the difficulty parameter. To efficiently estimate the regression coefficients of the predictors, locally D-optimal designs are developed. After an introduction to the Rasch Poisson counts model and the Rasch Poisson-Gamma counts model we will specify these models as a particular generalized linear mixed model. Based on this embedding optimal designs for both models including several binary explanatory variables will be presented. Therefore, we will derive conditions on the effect sizes of certain designs to be locally D-optimal. Finally, it is pointed out that the results derived for the Rasch Poisson models can be applied for more general Poisson regression models which should receive more attention in future psychological research.
\end{abstract}
\vspace{0.5cm}

\noindent
\textbf{Key Words:} optimal design, Poisson Rasch counts model, Poisson-Gamma Rasch counts model, item response theory 
%
\vspace{3mm}

\section{Introduction}
The Rasch Poisson counts model (RPCM) is the first Rasch model developed by the Danish mathematician Rasch (1960). It is the counterpart to the logistic Rasch model published some times later (Rasch, 1966a, 1966b). Both models predict the probability of responses by two parameters, item difficulty (easiness) and person ability. While the logistic Rasch model assumes binomially distributed responses linked by a logistic function to the difference of both parameters, the RPCM is based on responses distributed according to a Possion distribution and a logarithmic link function. Thus, the logarithm of the expectation of Poisson distributed scores consists of the sum of item (difficulty) and a person (ability) parameter (see details below).

Both models share the appealing feature that Rasch (1966, 104f) called "specific objectivity": "The comparison of any two subjects can be carried out in such a way that no parameters are involved other than those of the two subjects ... Similarly, any two stimuli can be compared independently of all other parameters than those of the two stimuli as well as the parameters of the subjects having been replaced with observable numbers. lt is suggested that comparisons carried out under such circumstances be designated as specific objective".

Unlike the logistic Rasch model, the RPCM has not got very much attention although many educational and psychological tests, especially those measuring human abilities, result in count data. Rasch (1960) used the RPCM for analyzing oral reading measured by the number of words read and the number of errors which turned out to be Poisson distributed. Other examples for tests scores measuring cognitive abilities which often follow a Poisson distribution include the main dimensions of human intelligence, mental speed, divergent thinking and memory. Using covariate-adjusted frequency plots, Holling, B{\"o}hning and B{\"o}hning (2015) proved that mental speed scores of a well-established German intelligence test follow a Poisson distribution (see also Doebler \& Holling, 2016), while Forthmann, Gerwig, Holling, Celik, Storme and Lubart (2016) provided evidence for this distributional assumption when analyzing scores for divergent thinking, e.g. number of ideas generated.  Extended Poisson models have been applied to further scores of established cognitive ability tests, such as the Kit of Reference Test for Cognitive Factors by ,e.g., Jansen and van Duijn (1992), Meredith (1968) Verhelst and Kamphuis (2009) or Ogasawara (1996). Finally, as a clinical example, the analysis of generalized anxiety orders by Poisson regression should be mentioned (Zainal \& Newmann, 2017).

In the logistic Rasch model as well as in the RPCM, person parameters are often specified as random effects. Choosing the Gamma distribution for the PRCM results in the so-called Rasch Poisson Gamma counts model (RPGCM). The Gamma distribution is the conjugate a-priori distribution for the Poisson distribution, leads compared to other distributions, e.g., normal or log-normal distribution, to favourable statistical properties of this model. The unconditional distribution of the responses is the generalized negative binomial distribution which allows for overdispersion (e.g., Molenberghs, Verbeke, Dem\'etrio \& Vieira, 2011).

An important extension of the logistic Rasch model is the well-known linear logistic model (LLTM) developed by Fisher (1973). This model has often been successfully applied for the development of educational and psychological tests. Here, the difficulty parameter is partly explained by item characteristics that correspond to certain cognitive operations that are required to correctly solve an item. Usually, such an item feature is represented by a binary variable indicating whether this item feature is present or not, i.\,e. whether the corresponding cognitive operation is required or not.

In the same way the RPCM and RPGCM can be extended, i.e., the item (difficulty/easiness) parameter consists of a sum of weighted binary predictors. These models allow for investigating the same research questions as can be investigated by the LLTM, but presupposes Poisson distributed scores. Such a research question refers to the calibration of rule-based generated items which are Poisson distributed as, e.g., developed by Doebler and Holling (2016).

To give an illustrative example we present numerical rule-based items for testing mental speed. These items show different levels of difficulty due to the rules, i.e., item features involved. These items consist of a set of e.\,g., 200 numbers (stimuli), such as 567, 1234, 2452, 1375,... The basic task is to mark all numbers with four digits. Now further item features can be specified in the instructions that has to be followed. An example of such rules determining the difficulty of an item family are: Mark those numbers which additionally satisfy the condition where the number is divisible by 2 (rule 1), greater than 600 (rule 2) has at least two identical digits (rule 3). Thus, an item family may be defined by the instruction to select all numbers which are divisible by 2 and less than 1200. Here, the sets of numbers are incidentals, that is, they are composed of numbers in such a way that they should have no impact on the difficulty.  The respondents have to work as fast as possible within a sufficiently short time interval such that it is hardly possible to work on all stimuli.

The calibration of such a test requires the extended Rasch Poisson model including binary predictors. Here, the main goal is to estimate the influence of the item features by the regressions coefficients of the binary predictors. An efficient calibration of such a test using the extended RPCM or RPGCM can best be accomplished by applying optimal design principles. In general, optimal design provides an important means in designing and calibrating tests. A recent overview of these issues is provided in the Handbook of Item Response Theory (van der Linden; 2016, 2017, 2018) by Berger (2018) and van der Linden (2018), while Holling and Schwabe (2016) contribute a chapter about statistical optimal design problems in psychometrics.

The optimal design problem to efficiently estimate the regression coefficients of the RPCM and RPGCM is to specify a design matrix according to a certain criterion, e.g., shortest lengths of confidence intervals for the regression coefficients. Given three different item features and eight different items, $\mathbf{D}_1$ and $\mathbf{D}_2$ would be examples for a design matrix. The rows represent the different items and the columns represent the intercept as well as the three item features. 0’s and 1’s indicate whether the corresponding item feature is present or not, respectively.

\[
\mathbf{D}_1=
\begin{tabular}{ccc}
$\left(
\begin{tabular}{rrrr}
$1$ & $1$ & $1$&  $1$  \\
$1$ & $1$ & $1$ & $0$ \\
$1$ & $1$ & $0$ & $1$ \\
$1$ & $1$ & $0$ & $0$ \\
$1$ & $0$ & $1$&  $1$  \\
$1$ & $0$ & $1$ & $0$ \\
$1$ & $0$ & $0$ & $1$ \\
$1$ & $0$ & $0$ & $0$
\end{tabular}
\right)$
$\mathbf{D}_2=$
$ \left(
\begin{tabular}{rrrrr}
$1$ & $1$ & $0$&  $0$  \\
$1$ & $0$ & $1$ & $0$ \\
$1$ & $0$ & $0$ & $1$ \\
$1$ & $0$ & $0$ & $0$ \\
$1$ & $1$ & $0$&  $0$  \\
$1$ & $0$ & $1$ & $0$ \\
$1$ & $0$ & $0$ & $1$ \\
$1$ & $0$ & $0$ & $0$
\end{tabular}
\right)$
\end{tabular}
\]

The first column of these design matrices indicates the intercept while the following ones represent three binary predictors (factors).
Which design should be used for a certain regression model with an intercept and three binary predictors? It is well-known that for members of the general linear model, design $\mathbf{D}_1$, the full factorial design, is D-optimal, since it minimizes the volume of the confidence ellipsoid of the four parameters for a given confidence level. Given this optimality criterion, design $\mathbf{D}_1$ should be chosen when the responses are normally distributed. The quality of both designs can be compared by computing the ratio of the volumes of the confidence ellipsoids for all parameters to be estimated. Given a linear model as ``true'' model, this ratio yields 2.7 when $\mathbf{D}_2$ is related to $\mathbf{D}_1$. Thus, the confidence ellisoid is considerably smaller for $\mathbf{D}_1$ than for $\mathbf{D}_2$.

For linear models, the D-optimality is independent of the true parameters and will additionally
be optimal with respect to the G-optimality criterion which minimizes the maximum variance of the predicted values. Furthermore, the D-optimal full factorial design will also be optimal for the A-optimality criterion, leading to the smallest standard errors or confidence intervals for the parameters on the average.

However, if the extended RPCM or PRGCM are the appropriate model for the above example, design matrix $\mathbf{D}_2$ will be often a better choice than $\mathbf{D}_1$ as will be shown below. To be more precise $\mathbf{D}_2$ will be a D-optimal design for a certain region of parameters.
In general, for nonlinear models, optimal designs depend on the true parameters. Therefore, optimal designs can only be developed for
certain subsets of parameters which are called locally optimal designs. Furthermore, for nonlinear models main optimality criteria, such as D- and G-optimality, do not coincide. For these reasons the development of optimal designs for Poisson regression models is much more demanding than those for linear models.

In the following, we will consider D-optimality since the minimization of the volume of confidence ellipsoids is a frequently desired criterion. Finally, most other criteria than D-optimality depend on the scaling of the independent variables, hence for example, dummy-coding may lead to other results than effect-coding for other optimality criteria than the D-criterion. Thus, D-optimal designs for the Rasch Poisson model as well as the Rasch Poisson-Gamma model with several binary predictors will be derived in this article.

D-optimal designs for the Rasch Poisson model as well as the Rasch Poisson-Gamma model with only two binary predictors have already been developed by Gra{\ss}hoff, Holling, and Schwabe (2013, 2016). However, models incorporating only two predictors are very restrictive since often the influence of more than two item features is of interest and it is more difficult to find optimal designs in higher dimensions. Previous results on two features could be generalized for deriving a Lemma, however a ``new'' Theorem for providing the optimal designs for the general case of more than 2 features has to be derived. These results allow for planning more complex models with dependent count variables. For example, rule-based generated items according to Rasch Poisson models usually will incorporate a lot of binary items features as studies with the LLTM have shown.

In the next section, the Rasch Poisson and Poisson-Gamma counts model and its extension by incorporating covariates will be introduced. We will then derive the results for the Poisson-Gamma model in the general case with $K \geq 3$ binary predictors and obtain corresponding results for the Poisson model as a special case, by letting the random effect tend to zero. The paper ends with a short discussion. Proofs are deferred to an appendix.

\section{The Rasch counts model}

For the number $Y$ of correct answers to a task the Rasch Poisson counts model assumes that $Y$ is Poisson distributed and its mean  $\mu_Y=\theta\sigma$ is specified by the product of the \textit{ability\/} $\theta$ of the respondent (\textit{person\/}) and the \textit{easiness\/} $\sigma$ of the task (\textit{item\/}) such that the mean $\mu_Y$ increases with both the ability and the easiness.

This model is extended by the incorporation of the influence of various item features on the easiness $\sigma$ of the item.
The dependence of $\sigma$ on the item features $\mathbf{x}$ is explained by a linear predictor $\mathbf{f}(\mathbf{x})^\top\bdgr{\beta}$ using the log link, $\log(\sigma)=\mathbf{f}(\mathbf{x})^\top\bdgr{\beta}$ and, hence, the number $Y$ of correct answers associated to items with features $\mathbf{x}$ is Poisson with mean $\mu_Y=\theta\exp(\mathbf{f}(\mathbf{x})^\top\bdgr{\beta})$.

More specifically, the explanatory variables $\mathbf{x}=(x_1,...,x_K)^\top$ specify $K$ possible item features, $\mathbf{f}=(f_1,...,f_p)^\top$ is a vector of known regression functions to describe the structural influence of the item features, and $\bdgr{\beta}$ is a $p$-dimensional vector of unknown parameters quantifying the strength of the influence of the item features.
As for a correct solution an operation related to a particular item feature $k$ may either be required or not, this item feature can be expressed by a binary dummy variable $x_k$, where $x_k=1$, if the $k$th operation is required, and $x_k=0$ otherwise.
The linear component $\mathbf{f}(\mathbf{x})^\top \bdgr{\beta}$ can then be specified by a $K$-way layout with binary explanatory variables $x_k=0,\ 1$, $k=1,...,K$.
We further assume that there are only direct effects of the item features, (i.\,e.\, no interactions occur between the item features).
The vector is given by $\mathbf{f}(x_1,...,x_K)=(1,x_1,...,x_K)^\top$ and, thus, there are $p=K+1$ parameters.
The parameter vector $\bdgr{\beta}=(\beta_0,\beta_1,...,\beta_K)^\top$ consists of a constant term $\beta_0$ and the $K$ effects $\beta_1,...,\beta_K$ of the $K$ item features on the easiness.
With this model assumptions the expected (mean) response equals
$
\mu_Y=\theta\exp(\beta_0+\sum_{k=1}^K \beta_k\, x_k)
$.
In the particular case that none of the $K$ item features is present (i.\,e.\, $x_k=0$ for all $K$ item features), a basic item is presented
which will be solved with mean response $\mu_0=\theta\exp(\beta_0)$.

Typically items will become more difficult, when cognitive operations are required to correctly solve the item, and the mean response will decrease.
Therefore the coefficients $\beta_k$ have to be negative -- or eventually equal to zero, if the cognitive operation does not influence the easiness of the items.
Therefore we will assume throughout $\beta_k\leq 0$ for the effects of the item features $k=1,...,K$.

In the following we focus on the calibration step for the test items, that is, our main interest is in the dependence of the easiness of the items features or cognitive operations.
Here we consider two different models in which we assume either that the abilities $\theta$ of the respondents are known beforehand or that the respondents are randomly selected from a homogeneous population.
The first assumption leads to a pure Poisson model.
In the latter case the ability can be considered as a random effect.
Usually, for the ability a Gamma distribution is assumed which leads to the Poisson-Gamma model, see e.\,g.\ Verhelst and Kamphuis (2009).

To be more specific the conditional distribution of the number $Y$ of correct \mbox{answers} given the ability $\Theta=\theta$ is Poisson with mean $\theta\sigma$, where the easiness $\sigma$ of the item is as specified above, and the ability $\Theta$ is Gamma distributed with shape parameter $a>0$ and scale parameter $b>0$.
The marginal probabilities of $Y$ can be obtained by integration of the joint density with respect to $\theta$.
The resulting Poisson-Gamma distribution of $Y$ is (generalized) negative binomial with success probability $1/(b\sigma)$ and (generalized) number of successes $a$.
Hence, $Y$ has expectation $\mathrm{E}(Y)=ab\sigma$ and variance $\mathrm{Var}(Y)=(1+b\sigma)\mathrm{E}(Y)$.
If the expectation $\mu=ab\sigma$ is held constant, the Poisson distribution is obtained as a limiting case with intensity (mean) $\mu$ when the scale parameter $b$ tends to $0$, which relates to the situation of known ability.
For simplicity we consider further on the situation that each person receives exactly one item in order to ensure independence of the observations.

\section{Information and design}
The impact of an experimental setting on the quality of the maximum likelihood estimator of the parameter vector $\bdgr{\beta}$ is measured by the Fisher information matrix because the asymptotic covariance of the maximum likelihood estimator for $\bdgr{\beta}$ is proportional to the inverse of the information matrix.
Hence, maximization of a real valued function of the information matrix, like the determinant means a maximization of the precision of the maximum likelihood estimator.

Similar to the situation in generalized linear models, the Fisher information based on one observation at the setting $\mathbf{x}$ for the explanatory variable is obtained in the form
$
\mathbf{M}(\mathbf{x};\bdgr{\beta}) = q(\mathbf{x};\bdgr{\beta})^{-1}\,\mathbf{f}(\mathbf{x})\, \mathbf{f}(\mathbf{x})^{\top}
$ with a weight function $q(\mathbf{x};\bdgr{\beta})^{-1}$ which measures to which amount an observation at item $\mathbf{x}$ contributes to the information (cf. Holling \& Schwabe, 2016).
The inverse weight function $
q(\mathbf{x};\bdgr{\beta})
=
(b+\exp(-\mathbf{f}(\mathbf{x})^{\top}\bdgr{\beta}))/(ab)
$
occurring in the information matrix depends on both the setting $\mathbf{x}$ (specifying the item features or cognitive operations) and the parameters $\bdgr{\beta}$.
For the pure Poisson case (limiting case for $b\to 0$ while keeping the mean $ab$ fixed) the inverse weight function can be obtained by setting $ab=\theta_0$ and then formally letting $b=0$, i.\,e.\,
$q(\mathbf{x};\bdgr{\beta})
=
\exp(-\mathbf{f}(\mathbf{x})^{\top}\bdgr{\beta})/\theta_0
$.

Since observations (answers to different items) are assumed to be statistically independent,
the normalized (per observation) information matrix equals
$
\mathbf{M}(\mathbf{x}_1,\ldots,\mathbf{x}_N;\bdgr{\beta})={\textstyle{\frac{1}{N}}}\, {\textstyle{\sum_{i=1}^N}}\, \mathbf{M}(\mathbf{x}_i;\bdgr{\beta})
$
for an exact design $(\mathbf{x}_1,\ldots,\mathbf{x}_N)$ with answers to $N$ items specified by their settings $\mathbf{x}_1,...,\mathbf{x}_N$ for the features.

Finding optimal settings $\mathbf{x}_1,...,\mathbf{x}_N$ is a discrete optimization problem.
Because the solution of such problems is difficult, we embed this optimization problem into the continuous setup of approximate designs (see Silvey 1980).
For this, note first that the information matrix $\mathbf{M}(\xi;\bdgr{\beta})$ may be rewritten as normalized information matrix
$
{\textstyle{\sum_{i=1}^n}}\, w_i\,\mathbf{M}(\mathbf{x}_i;\bdgr{\beta})
$,
where now $\mathbf{x}_1,...,\mathbf{x}_n$ denote mutually different settings occurring in the design, $w_i=N_i/N$ are the corresponding proportions and $N_i$ equals the numbers of replications for $\mathbf{x}_i$, $i=1,...,n$, within the design.

For an approximate design a relaxation is introduced which allows continuous weights $w_i$ instead of being multiples of $1/N$.
Thus an approximate design $\xi$ is defined by a set of mutually different settings $\mathbf{x}_1,\ldots,\mathbf{x}_n$ and corresponding weights $w_1,\ldots,w_n\geq 0$ satisfying $\sum_{i=1}^n w_i=1$.
This approach can be adopted here, as typically the number $N$ of items presented is large and optimal or, at least, efficient exact designs can be obtained by rounding the weights.
The information matrix for an approximate design is defined by
$
\mathbf{M}(\xi;\bdgr{\beta}) =  {\textstyle{\sum_{i=1}^n}}\, w_i\, q(\mathbf{x}_i;\bdgr{\beta})^{-1}\, \mathbf{f}(\mathbf{x}_i)\, \mathbf{f}(\mathbf{x}_i)^{\top}
$,
which coincides with the normalized information matrix, when $\xi$ is an exact design.

The aim of an optimal design is now to maximize the information matrix.
In general, this cannot be achieved simultaneously for the whole information matrix.
Therefore a real valued (one-dimensional) function of the information matrix will be optimized instead.
In the literature there is a broad choice for meaningful functionals (see e.\,g.\ Holling \& Schwabe, 2016).
We will adopt here the most popular criterion of $D$-optimality which aims at minimizing the volume of the (asymptotic) confidence ellipsoid for estimating the parameters $\bdgr{\beta}$ and has the advantage to be invariant with respect to rescaling and relabeling.
From the definition of the information matrix it is apparent that in contrast to linear models the information matrix and, hence, the optimal design will depend on the parameter $\bdgr{\beta}$.
Taking this into account a design $\xi^*$ will be called locally $D$-optimal at $\bdgr{\beta}$ if it maximizes the determinant $\det(\mathbf{M}(\xi;\bdgr{\beta}))$ of the information matrix.

In the present situation of a Poisson-Gamma model the information
$\mathbf{M}(\xi;\bdgr{\beta}) =\theta_0\exp(\beta_0)\mathbf{M}_0(\xi;\bdgr{\beta})$
is proportional to the mean ability $\theta_0=ab$ and to the easiness $\exp(\beta_0)$ of the basic item, where
$\mathbf{M}_0(\xi;\bdgr{\beta})$ denotes the information matrix in the standardized case $\theta_0=1$ and $\beta_0=0$.
Thus the determinant $\det(\mathbf{M}_0(\xi;\bdgr{\beta}))$ of the information matrix in the standardized case can be optimized independently of $\theta_0$ and $\beta_0$ and we will restrict ourselves to this standardized case ($\theta_0=1$, $\beta_0=0$) without loss of generality throughout the remainder of the paper.

This restriction is also applicable to the pure Poisson case when the ability of the respondent is assumed to be known.
If in that case the persons can be deliberately chosen, that is, any member of the population can be determined to participate in a certain study, since persons with the highest ability provide the most information. Mean and variance are equal for Poisson distributed variables, hence, the variation coefficient decreases with increasing mean, that is, the information
increases with growing mean. The weight function $q^{-1}$ is proportional to the mean, hence, growing with increasing means. This is also true for the Poisson Gamma-model.

\section{Optimal design for the Rasch counts model with binary predictors}
\label{two-way}

The case of $K=2$ features has been considered by Gra{\ss}hoff et al. (2013) for the pure Poisson counts model and by Gra{\ss}hoff et al. (2016) for the Poisson-Gamma model.
They established that in the case of sufficiently large negative values of the regression coefficients $\beta_1$ and $\beta_2$ the optimal design $\xi^* = \xi_{\mathbf{0}}$ avoids the most difficult item $\mathbf{x}=(1,1)$ in which both item features are given.
In particular, they characterize a whole parameter region for which the design $\xi_0$ is uniformly optimal by means of a nonlinear inequality involving the 2 parameters $\beta_1$ and $\beta_2$ associated with the 2 features.
For $K\geq 3$ features their approach leads to a whole set of nonlinear inequalities involving up to all of the $K$ feature parameters (see Lemma 1 below) which, in general, are hard to check simultaneously.
This problem is then resolved by the subsequent Theorem 1 which allows the reduction to the simpler inequalities which involve only two of the parameters each.
The proof of Theorem 1 needs a couple of analytical tools and is given in the Appendix.
To be more precise, in the situation where up to $K\geq 3$ features can be involved there are $2^K$ different items, which can be presented.
The corresponding $2^K$ settings may be regarded as the vertices of a $K$-dimensional (hyper-)cube (for $K=3$ see Figure~1).
\begin{figure}[h]
\begin{center}
\includegraphics[scale=0.8]{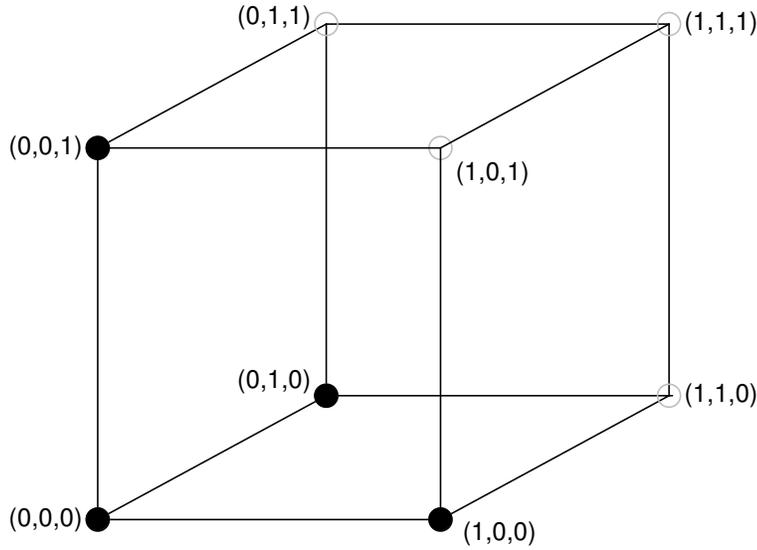}
\begin{minipage}{\captionwidth}
{\footnotesize{%
\caption{Graphical representation of the design points for a regression with three binary predictors. Points of the design $\xi_{\mathbf{0}}$ are highlighted.}
}}
\end{minipage}
\end{center}
\end{figure}

Optimality conditions have to be checked for each of the $2^K$ settings.
In the following we establish simple conditions for the local $D$-optimality of the design $\xi_{\mathbf{0}}$ with equal weights $w^*=1/(K+1)$ are assigned to those items, in which, at most, one of the item features is given, that is the basic item $\mathbf{x}_0=(0,0,...,0)$ and the $K$ ``one-feature'' items $\mathbf{x}_1=(1,0,...,0)$, $\mathbf{x}_2=(0,1,...,0)$, ..., $\mathbf{x}_K=(0,...,0,1)$, in which only one of the entries $x_k$ is equal to one (``feature $k$ is given'') and all other entries are equal to zero.
For the special case $K=3$, these settings are highlighted in Figure~1.

In Lemma~1 we first present a set of conditions on the inverse weight functions which are valid in general and do not make use of the particular structure of the Poisson-Gamma model. For brevity let $q_k(\bdgr{\beta})=q(\mathbf{x}_k;\bdgr{\beta})$ denote the inverse weight function at the setting $\mathbf{x}_k$ of the design $\xi_{\mathbf{0}}$ and $|\mathbf{x}|=\sum_{k=1}^K x_k$ the number of item features given in $\mathbf{x}$.
\newpage
\noindent
\textbf{Lemma~1. }
\textit{
The design $\xi_\mathbf{0}$ is $D$-optimal if the condition
$
(|\mathbf{x}|-1)^2 q_0(\bdgr{\beta})
+ {\textstyle{\sum_{k=1}^K}} x_{k} q_k(\bdgr{\beta})
\leq q(\mathbf{x};\bdgr{\beta})
$
is satisfied for all binary $\mathbf{x}$ with $|\mathbf{x}| \geq 2$.}
\vspace{5mm}

Without further assumptions it is not clear whether there exist parameter values for which all conditions in Lemma~1 are satisfied simultaneously.
Therefore we have to make use of the structure of the inverse weight functions in the underlying Poisson-Gamma model.
Note that in the Poisson-Gamma model the inverse weight functions $q(\mathbf{x};\bdgr{\beta})=b+\prod_{k=1}^{K}x_k\exp(-\beta_k)$ at the design points are given by $q_0=q_0(\bdgr{\beta})=b+1$ for the basic item and $q_k(\beta_k)=q_k(\bdgr{\beta})=b+\exp(-\beta_k)$, when only feature $k$ is given.
As we see in Theorem~1, for the characterization of the weight functions or, equivalently, the parameter values for which $\xi_{\mathbf{0}}$ is optimal we only require the weight functions for one or two item characteristics.
For this denote by $q_{jk}(\beta_j,\beta_k)=b+\exp(-\beta_j)\exp(-\beta_k)$ the inverse weight functions for an item in which features $j$ and $k$ are given.
\vspace{5mm}

\noindent
\textbf{Theorem~1. }
{\it In the Poisson-Gamma Rasch counts model with $K$ binary predictors and regression coefficients $\beta_k\leq 0$, the
design $\xi_\mathbf{0}$ is locally $D$-optimal at $\bdgr{\beta}$, if and only if the conditions
\begin{equation}
q_0(\bdgr{\beta}) + q_j (\bdgr{\beta}) + q_k(\bdgr{\beta}) \leq q_{jk}(\bdgr{\beta})
\label{cond-K-way}
\end{equation}
are satisfied for all pairs of features $1\leq j<k\leq K$.}
\vspace{5mm}

In fact, Theorem~1 establishes that the conditions in Lemma 1 are only to be proved for settings $\mathbf{x}$ with two non-zero entries ($|\mathbf{x}|=2$). By letting the variance of the gamma distribution tend to zero for the ability $\theta$ (or equivalently by formally setting $b=0$ in the proof of Theorem~1) we obtain the corresponding result for the pure Poisson Rasch counts model.

The two-dimensional conditions required in Theorem~1 coincide with the corresponding conditions for $K=2$.
Therefore, condition~(\ref{cond-K-way}) is satisfied for coefficients $\beta_k$ which have sufficiently large negative values, that is, a sufficiently strong decrease $\exp(\beta_k)$ in easiness or, equivalently, a sufficiently strong increase $\exp(-\beta_k)$ in difficulty $1/\sigma$.

\begin{figure}[h]
\begin{center}
\includegraphics[scale=0.8]{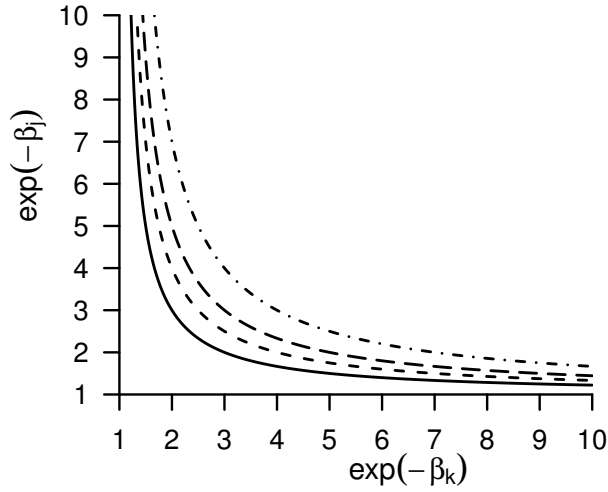}
\begin{minipage}{\captionwidth}
{\footnotesize{%
\caption{Graphical representation of the optimality condition. The regions, where $\xi_{\mathbf{0}}$ is locally $D$-optimal, dependent on the difficulty $\exp(-\beta_j)$ and $\exp(-\beta_k)$, are located above the curves for $b=2$ (dashes and dots), $b=1$ (long dashes), $b=0.5$ (short dashes), and $b=0$ (Poisson; solid line).}
}}
\end{minipage}
\end{center}
\end{figure}

While Lemma 1 is a generalization of previous results on $K=2$ features, Theorem 1 provides new and unexpected findings: Only conditions on the parameters of two features simultaneously are required. These conditions coincide with those imposed for less features and make the results applicable. Moreover the previous results solely rely on the general form of the information matrix while Theorem 1 makes use of the particular structure of the inverse weight function and thus introduces new concepts. The result is in so far unique that it cannot be extended to models with interactions: Kahle, Oelbermann, and Schwabe (2016) provided a counterexample to the conjecture that only conditions on up to 4 factors (features) would be required in models with first-order (two-factor) interactions.

Condition~(\ref{cond-K-way}) can be rephrased as
$
\exp(\beta_j)\leq (1-\exp(\beta_k))/(1+(1+2b)\,\exp(\beta_k))
$
in terms of the (relative decrease in) easiness or, equivalently, as
$
\exp(-\beta_j)\geq (\exp(-\beta_k)+1+2b)/ (\exp(-\beta_k)-1)
$
in terms of the (relative increase in) difficulty.

Figure~2 displays the dependence of the optimal design on the relative increase $\exp(-\beta_j)$ and $\exp(-\beta_k)$ in difficulty when features $j$ and $k$ are present.
If for all pairs of features the increase in difficulty lies in the region above the separating curves the
design $\xi_{\mathbf{0}}$ is locally $D$-optimal.

For smaller values of the difficulty (below the curves) also items have to be presented with more than one feature.
The corresponding weights have to be determined numerically.
It should also be noted that in the case that the cognitive operations do not influence on the easiness ($\beta_k=0$) all items are equally difficult, and the full factorial design which presents all $2^K$ items with the same weight $w^*=1/2^K$ is optimal.

\section{Discussion}

In this article we considered locally $D$-optimal designs for the Rasch Poisson counts
model and Poisson-Gamma model including binary explanatory variables. If additional cognitive operations are required to solve a certain set of items,
the number of correctly solved items usually does not increase and all regression coefficients are less or equal 0.
In this case, a
design design $\xi_{\mathbf{0}}$ consisting of points which incorporate at most one item feature proved to be optimal when the size of the regression coefficients is large.
Although these designs are only locally optimal they retain their local optimality for a wide range of parameter values.
Under the restriction of non-positive effects ($\beta_k\leq 0$ for all item features, $k=1,...,K$) these
designs attain a minimal value of $2^{(K+2)/(K+1)}/(K+1)$ for the efficiency at indifference ($\beta_1=\ldots=\beta_K=0$). Here, efficiency is defined as
$\det(M(\xi;\bdgr{\beta})/\det(M(\xi^*_{\bdgr{\beta}};\bdgr{\beta}))^{{1/p}}$ with
$\xi^*_{\bdgr{\beta}}$ as the locally optimal design. Hence, the efficiency measures the amount of observations to be made, when the optimal design $\xi_{\bdgr{\beta}^*}$
is used, to obtain the same information as under the design $\xi$.

In particular, for $K=2$ features the efficiency of the design $\xi_{\mathbf{0}}$ is about $84$\% (see Gra{\ss}hoff et al., 2013), for $K=3$ it is $59$\%, and for $K=6$ it is still $31$\%.
On the opposite the standard full factorial $2^K$ design has a minimum value of $(K+1)/2^K$ for large negative values of $\beta_k$.
In particular, for $K=2$ features the efficiency of the full factorial design is $75$\% (see Gra{\ss}hoff et al., 2013), for $K=3$ it is $50$\%, and it decreases rapidly when the number of features is increasing ($11$\% for $K=6$).
Hence, for a larger number of features there should be interest in designs which are less sensitive to changes in $\bdgr{\beta}$ like maximin efficient or weighted (``Bayesian'') designs when there are doubts about the magnitude of the effects.
This will be a topic of further research.
If one or more features do not have any influence on the easiness ($\beta_k=0$), then optimal designs can be generated as a product-type design in which equal weights are assigned to the combinations of all features without influence given the settings of the features with influence, and in the optimal marginal design on the subset of features with influence the weights have to be determined numerically (see Gra{\ss}hoff et al., 2013 and 2016, for the case $K=2$).
The local optimality of the
proposed designs is in accordance with the results of Schmidt and Schwabe (2017) who derived optimal designs in a Poisson-Gamma regression model with multiple continuous predictors.
In fact, if we restrict their continuous design region to the present case of binary predictors, their optimal design happens to coincide with the design obtained here for particular parameter values $\beta_1=...=\beta_K$ which are equal to $-2$ in the special case of Poisson regression.
In a related approach binary regressors have also been studied by Yang, Mandal, and Majumdar (2012) to characterize optimal designs in the case of binary response instead of count data using a logistic regression model.

For the derived optimal designs it has to be noted that a sufficient condition for the optimality of the proposed designs is satisfied when the application of any cognitive operation decreases the easiness at least by the factor 0.414.
Such a factor might be unrealistic for some tests such as mental speed tests involving very elementary cognitive operations.
However, when tests, e\,g. reasoning tests, require more complex operations, the proposed
designs will be locally $D$-optimal if the cognitive operations are sufficiently difficult, that is, easiness is decreased by the factor 0.414 compared to the basic item.

The extended RPCM outlined above includes binary predictors for the easiness parameter $\sigma$.
Due to the symmetry of the item parameter $\sigma$ and the ability parameter $\theta$ in the RPCM the derived optimal designs are also valid for models with binary explanatory variables for the person parameter $\theta$.
The resulting optimal test designs may be used, for example, to analyze the impact of variables such as gender on the ability.
Furthermore, the optimal designs derived above are also valid for models including binary predictors for the person as well as the item parameter.

Finally, it has to be mentioned that the optimal designs developed in this article do not account for interaction effects of the predictors.
But, such interactions may occur as can be derived from empirical studies where significant interaction effects between item features were found for tests measuring reasoning (e.\,g. Bertling, 2014).
Motivated by the present work, Kahle et al. (2016) show for the pure Poisson case by algebraic methods that also in the case of complete interactions up to degree $d$ designs which consist of items in which at most $d$ item features are given turn out to be optimal if the parameter effects are synergetic and sufficiently strong. They can provide a characterization by a set of nonlinear inequalities in all parameters similar to the present Lemma 1. But, they also show that a reduction of the conditions similar to Theorem~1 does not hold for interactions in general.
Thus, the derivation of optimal designs including interaction effects is of further
interest.

Last but not least, we would like to point out that the results derived above can be applied to the design of experiments when responses are Poisson distributed. Instead of linear regression models Poisson regression models may then be appropriate. Unlike the the RPCM and RPGCM, these models do not include person parameters. Since person parameters can be considered as nuisance parameters the derived results can also be applied for optimal designs for experiments as given for the PRCM.
Assume as an example an experiment (see Forthmann et al., 2016) for studying the number of ideas dependent upon three binary factors: (1) kind of instruction (generate creative ideas vs. create as many ideas as possible), (2) kind of stimulus (abstract vs. concrete) and (3) training (yes vs no). In this case $\mathbf{D}_2$ would be a D-optimal design for a wide region of true parameters, as has been proven above. These designs are also D-optimal for Poisson regression models including person parameters as fixed effects ,e.g., when repeated measurements are given. When person parameters are assumed to be Gamma distributed the results derived for RPGCM are to be considered.

A study by Vives, Losilla, and Rodrigo (2006) shows that experimental design for Poisson regression models is an important subject in psychological research. These authors reviewed a random sample of articles published between 2002 and 2006 in psychological journals with the highest impact factor, such as Biological Psychology or Journal of Personality and Social Psychology. The selection of those articles with response variables predicted by several independent variables yielded a sample of 457 regression models. The dependent variable of 40\% of these models was a count variable and the expectations of more than a third of these models were relatively low. Instead of using linear models (e.g. analysis of variance, multiple regression) Poisson regression models would be often appropriate to analyse such data. Thus, further development of optimal design for Poisson regression models is an important task for the future.


\section*{Appendix: Proofs}

Throughout this section we suppress the dependence on $\bdgr{\beta}$ for notational ease.

\vspace{5mm}
\noindent
\textbf{Proof of Lemma~1}

Let $\mathbf{F}_{\mathbf{0}}=(\mathbf{f}(\mathbf{x}_0),...,\mathbf{f}(\mathbf{x}_K))^\top$ be the design matrix and $\mathbf{Q}_{\mathbf{0}}=\mathrm{diag}(q_0,q_1,...,q_K)$ the diagonal matrix of the corresponding inverse weight functions for $\xi_{\mathbf{0}}$ .
With this notation the information matrix factorizes,
$
\mathbf{M}(\xi_{\mathbf{0}})= \mathbf{F}_{\mathbf{0}}^{\top} \mathbf{Q}_{\mathbf{0}}^{-1} \mathbf{F}_{\mathbf{0}}/(K+1)
$.
As $\mathbf{F}_{\mathbf{0}}$ and $\mathbf{Q}_{\mathbf{0}}$ are square matrices also the inverse
of the information matrix factorizes.
Hence, the sensitivity function
$
d(\mathbf{x};\xi_{\mathbf{0}})
 =
(K+1)\, 	q(\mathbf{x})^{-1} \mathbf{f}(\mathbf{x})^{\top} \mathbf{M}(\xi_{\mathbf{0}})^{-1}\mathbf{f}(\mathbf{x})
$
also factorizes to
$
d(\mathbf{x};\xi_{\mathbf{0}})
 =
(K+1)\, 	q(\mathbf{x})^{-1} (\mathbf{F}_{\mathbf{0}}^{-\top}\mathbf{f}(\mathbf{x}))^{\top} \mathbf{Q}_{\mathbf{0}} \mathbf{F}_{\mathbf{0}}^{-\top} \mathbf{f}(\mathbf{x})
$,
where
$
\mathbf{F}_{\mathbf{0}}^{-\top}
$
is the inverse of the transpose of the design matrix.

By $\mathbf{F}_{\mathbf{0}}^{-\top} \mathbf{f}(\mathbf{x}) =(1-|{\mathbf{x}}|,\mathbf{x}^{\top})^{\top}$ the sensitivity function simplifies to
\[
d(\mathbf{x};\xi_{\mathbf{0}})
= (K+1)\lambda(\mathbf{x})\left((|\mathbf{x}|-1)^2/\lambda_0+{\textstyle{\sum_{k=1}^K}} x_{k}^2/\lambda_k\right) \,.
\]
Note that $x_k^2=x_k$ for $x_k=0$, $1$, and that $d(\mathbf{x};\xi_{\mathbf{0}})=K+1$ at the support points $\mathbf{x}_k$ of $\xi_{\mathbf{0}}$, (i.\,e.\, for $|\mathbf{x}| \leq 1$).
Hence, the condition of Lemma~1 implies that the sensitivity function is bounded by the number of parameters $p=K+1$,
which proves the $D$-optimality of $\xi_{\mathbf{0}}$ in view of
the celebrated Kiefer-Wolfowitz equivalence theorem (see Silvey, 1980).
\hfill$\Box$
\vspace{5mm}

\noindent
\textbf{Proof of Theorem~1}

As noted before $d(\mathbf{x}_i;\xi_{\mathbf{0}})=K+1$ for $|\mathbf{x}| \leq 1$.
For $|\mathbf{x}|=2$ the conditions of Theorem~1 and Lemma~1 coincide.
Hence it remains to prove that condition~(\ref{cond-K-way}) of Theorem~1 implies that the condition of Lemma~1 is satisfied for all $|\mathbf{x}|\geq 3$.

For $\mathbf{x}$ with $|\mathbf{x}|=m$ let $k_1,...k_m$ be those indices for which $x_{k_j}=1$.
Then in the present Poisson-Gamma model $q(\mathbf{x})=b+{\textstyle{\prod_{j=1}^m}} \exp(-\beta_{k_j})$, where $\exp(-\beta_k)\geq 1$.
For technical purposes we introduce the functions
$
\varphi_m(z_1,...,z_m) = (m-1)^2(b+1)+{\textstyle{\sum_{j=1}^m}} (b+z_{j})-(b+{\textstyle{\prod_{j=1}^m}} z_{j})
$
for $z_j\geq 1$ and $m\geq 2$.
We can rephrase the condition of Lemma~1 as
$
\varphi_m(\exp(-\beta_{k_1}),...,\exp(-\beta_{k_m}))\leq 0
$.
Hence, it is sufficient to show that $\varphi_m(z_1,...,z_m)\leq 0$ for all $z_1,...,z_m$ such that $\varphi_2(z_j,z_k)\leq 0$ for all pairs $z_j$, $z_k\geq 1$.
This can be proved recursively.
The comparison of $\varphi_{m+1}$ and $\varphi_m$ yields
\[
\varphi_{m+1}(z_1,...,z_{m+1})-\varphi_m(z_1,...,z_m)=2m(b+1)-({\textstyle{\prod_{j=1}^m}}z_j -1)(z_{m+1}-1) \,.
\]
It is easy to show that ${\textstyle{\prod_{j=1}^m}}z_j \geq 1+{\textstyle{\sum_{j=1}^m}} (z_{j}-1)$ for $z_j \geq 1$.
Hence, the right hand side of the above equation is bounded above by
$
{\textstyle{\sum_{j=1}^m}}\varphi_2(z_j,z_{m+1}) \leq 0
$
by assumption.
This proves $\varphi_{m+1}(z_1,...,z_{m+1})\leq\varphi_m(z_1,...,z_m)\leq 0$ by induction.
\hfill$\Box$

\vspace{5mm}
\noindent
{\bf Acknowledgement} This work was supported by grant Ho1286-6/Schw531-15 of the Deutsche Forschungsgemeinschaft.


\noindent
\section*{References}
\begin{description}
\item Berger, M. P. F. (2018). Item-Calibration Designs. In W. J. van der Linden (Ed.), {\it Handbook of item response theory: Vol. 3. Applications}. Boca Raton: Chapman \& Hall/CRC.
\item Bertling, J. P. (2014). Measuring reasoning ability: Applications of rule-based item generation. Doctoral dissertation, University of M\"unster, Germany. Retrieved from:
https://repositorium.uni-muenster.de/document/miami/9a4fcdd9-3ae3-4e7f-8270-cfc043d37054/diss\_bertling.pdf
\item Doebler, A., \& Holling, H. (2016). A processing speed test based on rule-based item generation and the Rasch Poisson Counts
model. \textit{Learning and Individual Differences, 52}, 121--128
\item Fischer, G. H. (1973). The linear logistic test model as an instrument in educational
research. {\it Acta Psychologica, 37}, 359--374.
\item Forthmann, B., Gerwig, A., Holling, H., Celik, P., Storme, M., \& Lubart, T. (2016). The Be-Creative Effect in Divergent Thinking: The Interplay of Instructions and Objekt Frequency. {\it Intelligence, 57}, 25--32.
\item Gra{\ss}hoff, U., Holling, H. \& Schwabe, R. (2013). Optimal design for count data with binary predictors in item response theory.
In: D. Uci\'{n}ski, A. C. Atkinson, \& M. Patan (Eds): {\it mODa10 - Advances in Model-Oriented Design and Analysis\/} (pp 117--124). Cham: Springer.
\item Gra{\ss}hoff, U., Holling, H. \& Schwabe, R. (2016). Optimal design for the Rasch Poisson-Gamma model.
In: J. Kunert, C. H. M\"uller, \& A. C. Atkinson (Eds): {\it mODa 11 - Advances in Model-Oriented Design and Analysis\/} (pp 133--141). Springer.
\item  Holling. H., B{\"o}hning, W., \& B{\"o}hning, D. (2015). The covariate-adjusted frequency plot for the Rasch Poisson counts model. {\it Thailand Statistician, 13}, 67--78.
\item Holling, H. \& Schwabe, R. (2016). Statistical optimal design theory. In: W. J. van der Linden (Ed.) (2016). {\it Handbook of item response theory: Vol. 2. Statistical tools} (pp 313--340). Boca Raton: Chapman \& Hall/CRC.
\item Jansen, M. \& van Duijn, M. (1992). Extensions of Rasch’s multiplicative Poisson model. {\it Psychometrika, 57}, 405-414.
\item Kahle, T., Oelbermann, K.-F., \& Schwabe, R. (2016). Algebraic geometry of Poisson regression. \textit{Journal of Algebraic Statistics\/}, \textit{7}, 29--44.
\item Meredith, W. M. (1968). The Poison distribution and Poisson process in psychometric theory. {\it Research Bulletin 68--42}, Educational Testing Service, Princeton, New Jersey.
\item  Molenberghs, G., Verbeke, G., Dem\'etrio, C.,\& Vieira, A. (2011). A Family of generalized linear models for repeated measures with normal and conjugate random effects. {\it Statistical Science, 25}, 325--347.
\item Ogasawara, H. (1996). Rasch’s multiplicative Poisson model with covariates. {\it Psychometrika, 61}, 73--92.
\item Rasch, G. (1960). {\it Probabilistic Models for Some Intelligence and Attainment Tests}. Danish Institute
for Educational Research, Copenhagen.
\item Rasch, G. (1966). An individualistic approach to item analysis. In P. Lazarsfeld and N. Henry (Eds.), {\it Readings in Mathematical Social Sciences}, pp. 89--107. Cambridge: MIT Press.
\item Schmidt, D. \& Schwabe, R. (2017). Optimal design for multiple regression with information driven by the linear predictor.
{\it Statistica Sinica, 27}, 1371--1384.
\item Silvey, D. (1980). {\it Optimal Design}. London: Chapman \& Hall.
\item van der Linden, W. J. (Ed.) (2016, 2017, 2018). Handbook of item response theory: Models, statistical tools, and applications (Vols.1-3). Boca Raton: Chapman \& Hall/CRC.
\item van der Linden, W. J. (2018). Optimal test design. In W. J. van der Linden (Ed.), {\it Handbook of item response theory: Vol. 3. Applications}. Boca Raton: Chapman \& Hall/CRC.
\item Verhelst, N. D. \& Kamphuis, F. H. (2009). A Poisson-Gamma model for speed tests. Measurement and
Research Department Reports 2009-2. Cito, Arnhem.
\item Vives, J., Losilla, J.-M., \& Rodrigo. M.-F. (2006). Count data in psychological research. {\it Psychological Reports, 98}, 821--835.
\item Yang, J., Mandal, A., \&  Majumdar, D. (2012). Optimal designs for two-level factorial experiments with binary response. {\it Statistica Sinica}, \textit{22},  885--907.
\item Zainal, N. \& Newman, M. (2017). Executive function and other cognitive deficits are distal risk factors of generalized anxiety disorder 9 years later. {\it Psychological Medicine}, 1-9.
\end{description}

\end{document}